# Inverse Confounding Analysis: Part 2. Applications to Real-World Observational Data

**Sergey Porotsky, PhD**

**E-mail: sp0912@hotmail.com**

***Abstract***. Unmeasured confounding factors in observational data may lead to biased estimates of the effect of an exposure on an outcome. Consequently, a central problem in causal inference based on observational data is sensitivity analysis with respect to unmeasured confounding. Existing sensitivity-analysis methods generally focus on worst-case bounds. We propose an exact method for quantifying the significance of confounding, defined here in terms of the complete range of analytical values of the stratification-based Risk Ratio over the set of all joint distributions compatible with the observed characteristics. We refer to the proposed method as **Inverse Confounding Analysis (ICA) -** it is described in [1].
This paper presents examples of applying ICA to several real-world statistical datasets and examines an important issue that has attracted criticism of the method: the practical feasibility of obtaining the input data required for its application. The results are compared with those of two other methods for confounding analysis, the E-value and the method proposed by Peña.

**Key words: Causal Inference; Confounding; Observational Study.**

## 1. Introduction

There are two fundamentally different approaches to collecting statistical data:

- Randomized Controlled Trials;
- Observational Data Collection.

Most statistical studies use the second approach, often referred to as passive data collection. This is the approach considered in this paper.

When E and D are binary variables, the Risk Ratio (RR) is commonly used to analyze the relationship between the exposure E and the outcome D. A high Risk Ratio between E and D does not, by itself, guarantee that E actually affects D. The Risk Ratio may be large even when there is no causal relationship between E and D. This phenomenon is commonly referred to as a “spurious correlation.”

Several methods can be used to assess whether an observed correlation (or a Risk Ratio, in the case of binary variables) is spurious. For discrete factors, stratification is among the most commonly used methods [2]. The basic idea is to divide the original cohort into sub-cohorts in which the factor U, suspected of causing the spurious correlation, is held constant. The data are then analyzed separately within each sub-cohort. If a spurious correlation is suspected, stratification may help identify the factor responsible for it, known as a confounder.

However, stratification (like other similar methods) can identify spurious correlation only with respect to factors that were measured during data collection. Neither stratification nor other similar methods solves the problem of detecting spurious correlations due to unmeasured factors. Several other methods have been proposed in recent years, including the E-value [3, 4] and the method proposed by Peña [5]. However, these methods also do not provide definitive estimates of the extent to which a potential confounder U distorts the relationship between the exposure E and the outcome D. This motivated the development of Inverse Confounding Analysis (ICA) method [1].

This preprint is a logical continuation of [1]. Whereas [1] focused primarily on mathematical derivations, the present preprint examines practical issues involved in applying the ICA method.

The remainder of this preprint is organized as follows:

- The E-value, Peña, and ICA methods are briefly reviewed.
- The requirements for determining the input data are described, and the characteristics of the input parameters used by the three methods are compared.
- The results obtained using the three methods are compared in confounding-analysis problems with both complete and incomplete information about potential confounders.

## 2. Brief Overview of the E-value, Peña, and ICA Methods

Let us introduce the following notation:

- RR_ED denotes the Risk Ratio between the exposure E and the outcome D, calculated by directly analyzing the original data, i.e., without accounting for the potential confounder:

  **RR_ED = P(D = 1 | E = 1) / P(D = 1 | E = 0).**

For definiteness and simplicity, we assume throughout that **RR_ED ≥ 1**. The same reasoning can be applied to the case RR_ED < 1 after appropriately relabeling the levels of the binary variables.

- RR_ED_strat denotes the Risk Ratio between E and D calculated while accounting for the potential confounder U by stratification followed by standardization to the marginal distribution of U:

  **RR_ED_strat = Up / Dn**, where (1)

  Up = P(U = 1)×P(D = 1 | E = 1, U = 1) + P(U = 0)×P(D = 1 | E = 1, U = 0), (2)

  Dn = P(U = 1)×P(D = 1 | E = 0, U = 1) + P(U = 0)×P(D = 1 | E = 0, U = 0). (3)

If complete data on all three factors E, U, and D were available, RR_ED_strat could readily be calculated. However, the complete joint data on E, U, and D are unavailable, and we therefore cannot perform full stratification by the values of the unmeasured potential confounder U (U = 1 and U = 0). Consequently, we must use approximate estimates of RR_ED_strat. We denote:

- **Low_RR_ED_strat** — the lower bound on the quantity RR_ED_strat;
- **Up_RR_ED_strat** — the upper bound on the quantity RR_ED_strat.

The main objective is to determine which of the following cases applies:

- RR_ED_strat ≤ 1: the potential confounder U fully explains the relationship between E and D; that is, the correlation between E and D is “spurious.”

or

- RR_ED_strat > 1: the potential confounder U cannot fully explain the relationship between E and D; thus, with respect to adjustment for U, the correlation between E and D is not “spurious.”

In practical studies, a threshold other than 1 (for example, 0.95 or even 0.9) may be used in the inequalities above. This does not change the essence of the methods.

## 2.1. The E-value Method

To estimate RR_ED_strat, the E-value method uses two additional input parameters (in addition to RR_ED):

- RR_EU_mod — a modified measure of association between E and the unmeasured factor U;
- RR_UD_mod — a modified measure of association between the unmeasured factor U and the outcome D.

For calculating RR_EU_mod, [4] proposes the following expression:

- RR_EU_mod = max{RR_EU_U1, RR_EU_U0}, where

RR_EU_U1 = P(U = 1 | E = 1) / P(U = 1 | E = 0);

RR_EU_U0 = P(U = 0 | E = 1) / P(U = 0 | E = 0).

For calculating RR_UD_mod, [4] proposes the following expressions:

- RR_UD_mod = max{RR_UD_E1, RR_UD_E0}, where (4)

RR_UD_E1 = max{P(D = 1 | E = 1, U = 1), P(D = 1 | E = 1, U = 0)} /
min{P(D = 1 | E = 1, U = 1), P(D = 1 | E = 1, U = 0)}, (5)

RR_UD_E0 = max{P(D = 1 | E = 0, U = 1), P(D = 1 | E = 0, U = 0)} /
min{P(D = 1 | E = 0, U = 1), P(D = 1 | E = 0, U = 0)}. (6)

Article [4] states and proves the following bound for RR_ED_strat:

- **RR_ED_strat ≥ Low_RR_ED_strat,** where

Low_RR_ED_strat = RR_ED/[(RR_EU_mod × RR_UD_mod)/(RR_EU_mod + RR_UD_mod − 1)].

Because the E-value method provides a one-sided bound on RR_ED_strat (only its lower bound, Low_RR_ED_strat), it cannot, in principle, establish that the potential confounder U fully explains the relationship between E and D, i.e., that RR_ED_strat ≤ 1.

## 2.2. The Peña Method

The method proposed by Peña uses five input parameters to estimate RR_ED_strat:

- pE = P(E = 1);

- pDE1 = P(D = 1, E = 1);
- pDE0 = P(D = 1, E = 0);
- m = min over e,u of P(D = 1 | E = e, U = u);
- M = max over e,u of P(D = 1 | E = e, U = u),

where e = 0, 1 and u = 0, 1.

Article [5] states and proves the following two-sided bounds for RR_ED_strat:

- **Low_RR_ED_strat ≤ RR_ED_strat ≤ Up_RR_ED_strat,** where

Low_RR_ED_strat = [pDE1 + (1 − pE) × m] / [pDE0 + pE × M],

Up_RR_ED_strat = [pDE1 + (1 − pE) × M] / [pDE0 + pE × m].

## 2.3. The ICA Method

To estimate RR_ED_strat, ICA uses five additional input parameters (in addition to RR_ED):

- pE = P(E = 1);
- pU = P(U = 1);
- pD = P(D = 1);
- RR_EU = P(U = 1 | E = 1) / P(U = 1 | E = 0), the Risk Ratio between E and the potential confounder U;
- RR_UD = P(D = 1 | U = 1) / P(D = 1 | U = 0), the Risk Ratio between the potential confounder U and the outcome D.

It is fundamentally impossible to determine the exact value of RR_ED_strat from these input parameters alone. Therefore, ICA analyzes RR_ED_strat value in two stages.

1. First, the lower and upper bounds of RR_ED_strat (Low_RR_ED_strat and Up_RR_ED_strat) are calculated. Article [1] provides simple analytical expressions for calculating these bounds from the input data listed above.

- If Up_RR_ED_strat ≤ 1, we can state with certainty that RR_ED_strat ≤ 1. Thus, the potential confounder U fully explains the relationship between E and D: the observed correlation is 100% “spurious.”
- If Low_RR_ED_strat > 1, we can state with certainty that RR_ED_strat > 1. Thus, the potential confounder U cannot fully explain the relationship between E and D.
- If Low_RR_ED_strat ≤ 1 while Up_RR_ED_strat > 1, the analysis of RR_ED_strat should be continued. In this case, the probability that RR_ED_strat ≤ 1 is determined, and this probability is used to assess how strongly the potential confounder U supports the explanation of the relationship between E and D.

For this purpose, the free parameter p111 = P(E = 1, U = 1, D = 1) is introduced. Article [1] derives an expression representing RR_ED_strat as a fractional-linear function of p111:

- **RR_ED_strat = (a + b × p111) / (c + d × p111),** (7)

where a, b, c, and d depend only on the input parameters pE, pU, pD, RR_ED, RR_EU, and RR_UD. Simple analytical expressions for the bounds of p111 (p111_low and p111_high) are given in [1]. The problem is that the probability distribution of p111 over the interval [p111_low, p111_high] is unknown, which prevents direct calculation of the required quantity P(RR_ED_strat ≤ 1).

In general, virtually nothing is known about the distribution of p111: not only are its characteristics (moments, mode, median, quantiles, etc.) unknown, but even qualitative information—such as whether values of p111 are more likely to lie near the upper rather than the lower end of the interval [p111_low, p111_high]—is unavailable. In this situation, the principle of maximum entropy [6] suggests that the most appropriate assumption is that p111 is uniformly distributed over interval [p111_low, p111_high]. For this case, [1] derives simple analytical expressions for calculating P(RR_ED_strat ≤ 1). If some information about the distribution of p111 is available (for example, its mode or a specified quantile), the maximum-entropy principle can also be applied subject to these constraints. Examples of such an approach in connection with the Cross-Entropy optimization method [7] are given in [8]. In this case, analytical expressions for the distribution of p111 and for P(RR_ED_strat ≤ 1) generally cannot be obtained; the results must be calculated only numerically.

## 3. Characteristics of the Required Input Data

Let us examine in greater detail the six input parameters required by ICA method (pE, pU, pD, RR_ED, RR_EU and RR_UD). Two natural questions arise:

- To what extent does the increase in the number of input parameters to six (compared with three parameters for the E-value method and five for the Peña method) make ICA difficult to apply in practice?
- Why can the free (unknown) parameter p111 not be determined in addition to the six input parameters listed above? If it could be determined, RR_ED_strat could be calculated exactly rather than through the probabilistic estimate P(RR_ED_strat ≤ 1).

Regarding the first question, the practical applicability of different confounding-analysis methods depends not so much on the number of input parameters as on the amount and type of information about the potential confounder that the analyst must obtain.

Accordingly, the difficulty of determining each input parameter can usefully be assessed in terms of two characteristics:

- Whether or not the parameter is related to the unmeasured potential confounder U;
- How many factors the parameter depends on.

The characteristics of the input parameters required by the E-value, Peña, and ICA methods are summarized in Table 1. The table shows that the input parameters have quite different properties. For ICA, for example, pE, pD, and RR_ED contain no information about the unmeasured potential confounder U. They describe the observed situation, and determining them is a routine part of analyzing the data for E and D. By contrast, pU, RR_EU, and RR_UD are related to the potential confounder U, making their determination more difficult. Similar considerations apply to the E-value and Peña methods:

- In the E-value method, RR_ED is not related to the potential confounder U, whereas RR_EU_mod and RR_UD_mod are.
- In the Peña method, pE, pDE1, and pDE0 are not related to U, whereas m and M are.

Therefore, when comparing the difficulty of determining the input parameters, it is useful to focus on those parameters that are related to the unmeasured potential confounder U.

| **Characteristic** | **E-value** | **Peña** | **ICA** |
|---|---|---|---|
| Total number of input parameters | 3 | 5 | 6 |
| Parameters related to U | 2 (RR_EU_mod, R_UD_mod) | 2 (m and M) | 3 (pU, RR_EU, RR_UD) |
| Of these: first-level parameters | 0 | 0 | 1 (pU) |
| Second-level parameters | 1 (RR_EU_mod) | 0 | 2 (RR_EU, RR_UD) |
| Third-level parameters | 1 (RR_UD_mod) | 2 (m and M) | 0 |

Table 1. Analysis of the input parameters used by the E-value, Peña, and ICA methods.

On the other hand, the difficulty of determining an input parameter depends substantially on the number of factors with which it is associated. For example, determining pU, which is related only to the confounder U, is considerably easier than determining other input parameters that are also related to U but additionally depend on one of the other factors: RR_EU, RR_UD, RR_EU_mod.

The greatest difficulty arises in determining input parameters that depend simultaneously on all three factors E, U, and D. These include RR_UD_mod in the E-value method, m and M in the Peña method. An examination of the expressions for RR_UD_mod, m, and M shows that their calculation requires knowledge of quantities such as $P(D = 1 \mid E = 1, U = 1)$, $P(D = 1 \mid E = 1, U = 0)$, and so on.

In essence, this requires stratification by the unmeasured factor U, which is very difficult to carry out in practice.

In some cases, the difficulty of determining such input parameters is addressed by replacing them with other parameters. For example, in study [11], the more accessible parameter RR_UD is used in the E-value calculations instead of the required third-level input parameter RR_UD_mod. RR_UD depends only on two factors (U and D) and does not require stratification by U for its determination. Naturally, this substitution introduces some error when RR_UD_mod is approximated by RR_UD in formulas (4) – (6), and consequently when the lower bound on RR_ED_strat is calculated. This issue is discussed in greater detail in Section 5.2 of [1].

The following conclusions can therefore be drawn:

- Unlike the E-value and Peña methods, ICA does not require direct determination of the conditional probabilities $P(D = 1 \mid E = e, U = u)$, which are, in essence, third-level input parameters.
- Consequently, in some practical situations, the type of information required by ICA may be substantially more accessible than the information required by the E-value and Peña methods.

This also clarifies the second question raised at the beginning of this section: why can the free (unknown) parameter p111 not be determined in addition to the six input parameters? This parameter is a third-level parameter, defined as $P(E = 1, U = 1, D = 1)$, and its determination essentially requires stratification by the unmeasured factor U. This is practically impossible. If such stratification were possible, we could directly and readily calculate the required value of RR_ED_strat using formulas (1) – (3).

## 4. Numerical Examples

This section illustrates the application of the E-value, Peña and ICA methods to three practical problems involving the search for confounders:

- Analysis of IVF statistics [9, 10] to assess the relationship between a woman's age and the success of In Vitro Fertilization. This example is notable because all relevant variables are known, allowing the value of RR_ED_strat for each potential confounder to be calculated exactly in advance using formulas (1) – (3), with stratification. This makes it possible to examine how accurately each method approximates RR_ED_strat.
- Analysis of meta-analysis data on the association between antidepressant use during pregnancy and the risk of miscarriage [11].
- Analysis of data on the relationship between sleep-onset time and cardiovascular disease [12].

### 4.1. Analysis of IVF Statistical Data

An initial analysis of IVF statistics for 2015–2016 [9, 10] produces a surprising result: women aged 45–50 appeared to have a higher rate of successful IVF outcomes than younger women aged 43–44. We use the following definitions:

- Factor E — "Patient Age": E = 0 for ages 43–44 years (5,630 observations), and E = 1 for ages 45–50 years (3,256 observations).
- Factor D — "IVF Success": D = 0 for an unsuccessful outcome (7,531 observations), and D = 1 for a successful outcome (1,355 observations).

Processing the data reported in [9] for 8,886 patients aged 43–50 yielded the following values:

- **pE = 0.366; pD = 0.152; pDE1 = 0.069; pDE0 = 0.084; RR_ED = 1.416.**

The value of RR_ED would seem to indicate that older women have a higher rate of successful IVF outcomes than younger women, which contradicts common expectations. It is therefore reasonable to suspect a spurious correlation between patient age and IVF success. Since the values of various factors are available for all 8,886 observations, an obvious way to search for a confounder is to stratify the data, one factor at a time, and determine which potential confounder yields to condition RR_ED_strat <= 1.

Let **RR_ED_strat_true** denote the exact value of RR_ED_strat for the confounder under consideration, calculated from formulas (1) – (3) when complete information on U is available. In fact, we have complete information on all 95 factors for each of the 8,886 patients aged 43–50.

Now suppose that the complete statistical data for U are unavailable, and that for each potential confounder we can calculate only the input parameters required by the E-value, Peña, and ICA methods. The results are summarized in Table 2.

**a) Potential confounder U: "Number of previous pregnancies".** U = 0 if the patient had no previous pregnancies (2,578 observations), and U = 1 if she had previous pregnancies (6,308 observations). From formulas (1) – (3), RR_ED_strat_true = 1.413.

The input parameters and output values for the three methods are as follows:

- E-value method: RR_EU_mod = 1.024 and RR_UD_mod = 1.656, yielding Low_RR_ED_strat = 1.403. Thus, RR_ED_strat > 1, and the potential confounder U cannot fully explain the positive relationship between E (patient age) and D (IVF success).

- Peña method: m = 0.090 and M = 0.193, yielding Low_RR_ED_strat = 0.814 and Up_RR_ED_strat = 1.634. It is therefore unclear whether the potential confounder U can fully explain the positive relationship between E and D or not.
- ICA method: pU = 0.710, RR_EU = 1.024, and RR_UD = 1.285, yielding Low_RR_ED_strat = 1.405, Up_RR_ED_strat = 1.427, and P(RR_ED_strat ≤ 1) = 0. Thus, RR_ED_strat > 1, and the potential confounder U cannot fully explain the positive relationship between E and D.

The E-value and ICA results both indicate that "Number of previous pregnancies" cannot fully explain the relationship between E and D. Therefore, the search for a confounder should continue.

**b) Potential confounder U: "Cause of Infertility — Male Factor".** U = 1 if the primary cause of infertility was classified as male factor (2,004 observations), and U = 0 otherwise (6,882 observations). From formulas (1) – (3), RR_ED_strat_true = 1.426.

The input parameters and output values for the three methods are as follows:

- E-value method: RR_EU_mod = 1.052 and RR_UD_mod = 1.185, yielding Low_RR_ED_strat = 1.405. Thus, RR_ED_strat > 1, and the potential confounder U cannot fully explain the positive relationship between E and D.
- Peña method: m = 0.127 and M = 0.213, yielding Low_RR_ED_strat = 0.919 and Up_RR_ED_strat = 1.564. It is therefore unclear whether the potential confounder U can fully explain the positive relationship between E and D or not.
- ICA method: pU = 0.226, RR_EU = 0.834, and RR_UD = 1.159, yielding Low_RR_ED_strat = 1.396, Up_RR_ED_strat = 1.468, and P(RR_ED_strat ≤ 1) = 0. Thus, RR_ED_strat > 1, and the potential confounder U cannot fully explain the positive relationship between E and D.

The E-value and ICA results indicate that "Cause of Infertility — Male Factor" also cannot fully explain the relationship between E and D. Therefore, the search for a confounder should continue.

**c) Potential confounder U: "Fresh Cycle".** U = 0 if fresh embryos were used in the cycle (6,308 observations), and U = 1 otherwise (2,578 observations). From formulas (1) – (3), RR_ED_strat_true = 1.360. The input parameters and output values for the three methods are as follows:

- E-value: RR_EU_mod = 1.426 and RR_UD_mod = 1.776, yielding Low_RR_ED_strat = 1.232. Thus, RR_ED_strat > 1, and this confounder cannot fully explain the positive relationship between E and D.

- Peña method: m = 0.111 and M = 0.214, yielding Low_RR_ED_strat = 0.857 and Up_RR_ED_strat = 1.639. It is therefore unclear whether this confounder can fully explain the positive relationship between E and D or not.
- ICA: pU = 0.290, RR_EU = 1.426, and RR_UD = 1.552, yielding Low_RR_ED_strat = 1.272, Up_RR_ED_strat = 1.429, and P(RR_ED_strat ≤ 1) = 0. Thus, this confounder cannot fully explain the positive relationship between E and D.

The E-value and ICA results indicate that "Fresh Cycle," like the previously considered factors, cannot fully explain the relationship between E and D. Therefore, the search for a confounder should continue.

**d) Potential confounder U: "Egg Source".** U = 0 if the patient's own egg was used (6,847 observations), and U = 1 if a donor egg was used (2,039 observations). From formulas (1) – (3), RR_ED_strat_true = 0.757.

The input parameters and output values for the three methods are as follows:

- E-value method: RR_EU_mod = 2.589 and RR_UD_mod = 9.011, yielding Low_RR_ED_strat = 0.643. It is therefore unclear whether RR_ED_strat ≤ 1 or RR_ED_strat > 1, i.e., whether this potential confounder can fully explain the positive relationship between E and D or not.
- Peña method: m = 0.034 and M = 0.084, yielding Low_RR_ED_strat = 0.419 and Up_RR_ED_strat = 3.079. It is likewise unclear whether this potential confounder can fully explain the positive relationship between E and D or not.
- ICA: pU = 0.342, RR_EU = 2.589, and RR_UD = 5.399, yielding Low_RR_ED_strat = 0.705, Up_RR_ED_strat = 1.008, and P(RR_ED_strat ≤ 1) = 0.966. Thus, this confounder is highly likely to fully explain the positive relationship between E and D. This conclusion is consistent with the value RR_ED_strat_true = 0.757, which is known when the complete data are available.

Thus, on the fourth attempt, ICA identified, with high probability, the confounder responsible for the spurious correlation between E and D. The E-value and Peña methods did not do so: their calculated values do not allow a reasonably clear conclusion as to whether "Egg Source" can fully explain the relationship between E and D.

The results in Table 2 support the following conclusions:

- ICA correctly identified all four cases in the sense that the value of RR_ED_strat_true fell within the calculated bounds.

- In the three cases where RR_ED_strat_true was greater than 1, ICA ruled out the possibility that the potential confounder under consideration fully explained the apparent relationship.
- For the potential confounder "Egg Source," ICA yielded 96.6% for the integrated estimate P(RR_ED_strat ≤ 1), which is consistent with the known value RR_ED_strat_true = 0.757 for this factor.

| **Potential confounder U** | **E-value : lower bound** | **Peña : lower bound** | **Peña: upper bound** | **ICA: lower bound** | **ICA: upper bound** | **ICA: P(RR_ED_strat ≤ 1)** | **Exact RR_ED_strat** |
|---|---|---|---|---|---|---|---|
| Previous pregnancies | 1.403 | 0.814 | 1.634 | 1.401 | 1.427 | 0 | 1.413 |
| Male Factor | 1.405 | 0.919 | 1.564 | 1.396 | 1.468 | 0 | 1.426 |
| Fresh Cycle | 1.232 | 0.857 | 1.639 | 1.272 | 1.429 | 0 | 1.360 |
| Egg Source | 0.643 | 0.419 | 3.079 | 0.705 | 1.008 | 0.966 | 0.757 |

Table 2. Results of the analysis of the IVF data.

### 4.2. Analysis of Antidepressant Use

Article [11] uses a meta-analysis to examine the association between antidepressant use during pregnancy and the risk of miscarriage. Based on the analysis of various datasets, [11] reports a Risk Ratio of RR_ED = 1.41 for the association between antidepressant use and miscarriage. The authors of [11] raise the natural question of whether this estimate could be due to a spurious correlation between E and D resulting from a confounder. Two potential confounders are considered: smoking and alcohol consumption during pregnancy. The variables are defined as follows:

- Factor E — use of antidepressants.
- Factor U — smoking or alcohol consumption.
- Factor D — risk of miscarriage.

The meta-analysis in [11] provides the following Risk Ratios:

- RR_EU = 10.25 for U = alcohol consumption, and RR_EU = 2.06 for U = smoking.
- RR_UD = 3.1 for U = alcohol consumption, and RR_UD = 1.32 for U = smoking.

It should be noted that the E-value calculations in [11] use the more accessible second-level parameter RR_UD instead of the third-level parameter RR_UD_mod required by [3, 4]. We likewise use RR_UD instead of RR_UD_mod in our calculations.

Based on information available on the Internet, the following approximate values were determined for the additional first-level input parameters:

- pE = 0.1; pD = 0.1; pU = 0.15, for both U = alcohol consumption and U = smoking.

It is not possible to determine the third-level input parameters pDE1 and pDE0 required by the Peña method from the Internet sources available to us. Therefore, the Peña method is not used for this example.

Also note that RR_EU > 1 for both alcohol consumption and smoking. Consequently, when using the E-value method, RR_EU_mod = RR_EU for both potential confounders. The results obtained using the E-value and ICA methods are summarized in Table 3.

| **Potential confounder U** | **E-value: lower bound** | **ICA: lower bound** | **ICA: upper bound** | **ICA: P(RR_ED_strat ≤ 1)** |
|---|---|---|---|---|
| Alcohol consumption | 0.55 | 0.23 | 4.47 | 0.165 |
| Smoking | 1.23 | 0.76 | 1.65 | 0.264 |

Table 3. Analysis of antidepressant-use data.

**a) Potential confounder U: alcohol consumption**. The output values are:

- E-value: Low_RR_ED_strat = 0.55. It is therefore unclear whether RR_ED_strat < 1 or RR_ED_strat > 1, i.e., whether the potential confounder fully explains the positive relationship between E and D or not.
- ICA: Low_RR_ED_strat = 0.23, Up_RR_ED_strat = 4.47, and P(RR_ED_strat ≤ 1) = 0.16. Thus, with probability 0.84 under the assumed uniform distribution of p111, this confounder does not fully explain the positive relationship between E and D.

**b) Potential confounder U: smoking**. The output values are:

- E-value: Low_RR_ED_strat = 1.23 > 1. Thus, the potential confounder U cannot fully explain the positive relationship between E and D.
- ICA: Low_RR_ED_strat = 0.76, Up_RR_ED_strat = 1.65, and P(RR_ED_strat ≤ 1) = 0.26. Thus, with probability 0.74 under the assumed uniform distribution of p111, this confounder does not fully explain the positive relationship between E and D.

It is interesting to examine why the conclusions based on the E-value and ICA calculations are broadly similar (the factor U cannot fully explain the positive relationship between E and D), yet differ substantially in their degree of certainty:

- The E-value calculations appear to guarantee that RR_ED_strat > 1.
- The ICA calculations indicate that RR_ED_strat > 1 only with a probability substantially below 100%.

The reason is that the key E-value inequality requires the input parameter RR_UD_mod to be calculated using formulas (4) – (6). However, study [11] uses the more accessible parameter RR_UD instead. In some situations, RR_UD may be smaller than RR_UD_mod, which can lead to an invalidly high value of Low_RR_ED_strat. This issue is discussed in greater detail in Section 5.2 of article [1].

### 4.3. Analysis of the Relationship between Sleep-Onset Time and Cardiovascular Disease

Article [12] analyzes the relationship between sleep-onset time and cardiovascular disease. To assess whether the observed association could be spurious, the authors performed analyses in strata defined by factors such as age, sex, obesity, cholesterol, hypertension, diabetes, and hypotension. However, data were not collected on factors such as "a person's temperament" or "watching evening television news." It is therefore possible that sleep-onset time and cardiovascular disease are not causally related, but instead share a common cause, such as temperament or watching television news in the evening.

It is therefore of interest to examine these two potential confounders as possible explanations for a spurious correlation. Unfortunately, the original data reported in [12] cannot be used directly to obtain the input parameters RR_ED, RR_EU and RR_UD, because [12] analyzes Hazard Ratios rather than Risk Ratios and uses Cox Proportional Hazards (CPH) survival models. Accordingly, we emphasize that the analysis below does not reproduce the study in [12]. It merely demonstrates how ICA method can be used to analyze hypothetical potential confounders for specified input values.

We consider binary variables and apply the E-value and ICA methods. The variables are defined as follows:

- Factor E — sleep-onset time: E = 0 for early sleep onset and E = 1 for late sleep onset.
- Factor U — potential confounder: (a) "Temperament": U = 1 for a strong temperament and U = 0 otherwise; (b) "Watching Evening News": U = 1 if the person watches the evening news and U = 0 otherwise.
- Factor D — cardiovascular disease: D = 1 if cardiovascular disease is present and D = 0 otherwise.

To obtain approximate values of RR_ED_strat for the potential confounders "Temperament" and "Watching Evening News," we use input data estimated from information available on the Internet. For the E-value method, RR_UD_mod cannot be determined from these sources because doing so would require a detailed analysis stratified by U. Instead, we use RR_UD as a first approximation. RR_UD is a second-level input parameter (it does not depend on E) and is therefore much easier to determine than the third-level parameter RR_UD_mod. The results are summarized in Table 4.

The Peña method cannot be used to obtain an approximate estimate of RR_ED_strat in this example because it requires the third-level input parameters m and M, which are practically impossible to determine from the Internet sources available to us.

The following approximate input values were obtained from Internet sources:

- pE = 0.75, pD = 0.04, RR_ED = 1.3.
- For the potential confounder “Temperament”: pU = 0.5, RR_EU = 1.4, RR_UD = 1.1.
- For the potential confounder “Watching Evening News”: pU = 0.2, RR_EU = 2.0, RR_UD = 10.0.

**a) Potential confounder U: “Temperament”.** The output values are:

- E-value: Low_RR_ED_strat = 1.27 > 1. The potential confounder U cannot fully explain the positive relationship between E and D.
- ICA: Low_RR_ED_strat = 1.02, Up_RR_ED_strat = 1.57, and P(RR_ED_strat ≤ 1) = 0. The potential confounder U likewise cannot fully explain the positive relationship between E and D.

**b) Potential confounder U: “Watching Evening News”.** The output values are:

- E-value: Low_RR_ED_strat = 0.72. It is therefore unclear whether RR_ED_strat ≤ 1 or RR_ED_strat > 1, i.e., whether the potential confounder U fully explains the positive relationship between E and D or not.
- ICA: Low_RR_ED_strat = 0.69, Up_RR_ED_strat = 1.28, and P(RR_ED_strat ≤ 1) = 0.677. Thus, under the assumed uniform distribution of p111, this confounder has a 67.7% probability of fully explaining the positive relationship between E and D.

Thus, among the two potential confounders considered, ICA identifies “Watching Evening News” as the more likely explanation for the spurious correlation between “sleep-onset time” and “cardiovascular disease” for the assumed input values.

| **Potential confounder U** | **E-value: lower bound** | **ICA: lower bound** | **ICA: upper bound** | **ICA: P(RR_ED_strat ≤ 1)** |
|---|---|---|---|---|
| Temperament | 1.27 | 1.02 | 1.57 | 0 |
| Watching Evening News | 0.72 | 0.69 | 1.28 | 0.677 |

Table 4. Analysis of the relationship between sleep-onset time and cardiovascular disease.

## 5. Conclusions

This paper examines the ICA (Inverse Confounding Analysis) method proposed in [1] and compares its capabilities with those of the E-value [3, 4] and Peña [5] methods with respect to:

- the determination of the required input data;

- the accuracy and informativeness of estimates of RR_ED_strat.

With respect to input-data requirements, the analysis suggests that although ICA uses more input parameters (six) than the E-value (three) and Peña (five) methods, the parameters required by ICA may be more feasible to determine in practice than those required by the E-value and Peña methods. With respect to the comparison of the methods, the examples considered show that ICA provides more informative estimates of RR_ED_strat than the E-value and Peña methods. In some cases, ICA substantially narrows the admissible interval for RR_ED_strat and provides an additional integrated estimate, P(RR_ED_strat ≤ 1).

I am grateful to Michael Feldman for helpful discussions and remarks. In preparing the manuscript, I used ChatGPT to assist with the translation into English and to help express the material more clearly. However, all mathematical content and reasoning are my own.